\documentstyle[preprint,aps]{revtex}
\input epsf
\newcommand{\beq}{\begin{equation}}
\newcommand{\eeq}{\end{equation}}
\input epsf
\epsfverbosetrue

\begin{document}

\draft

\title{Vertex corrections in antiferromagnetic spin fluctuation theories}

\author{M.H. Sharifzadeh Amin and P.C.E. Stamp}

\address{ Physics Department
and Canadian Institute for Advanced Research,
 University of British Columbia, 6224 Agricultural
Rd., \\  Vancouver, BC, Canada V6T 1Z1}

\maketitle

\begin{abstract}
We calculate the first vertex correction  $\delta\Lambda_{\bf k}$ to
 the bare vertex $\Lambda_\circ$ in the nearly antiferromagnetic spin fluctuation Fermi
 liquid theory of the cuprate superconductor ${\rm YBa_2Cu_3O_7}$. It is
 calculated for $\bf k$ on the Fermi surface, and
 ${\bf Q}=(\pm {\pi\over a}, \pm {\pi\over a})$. We find that the dimensionless ratio
 $|\delta\Lambda_{\bf k}|/\Lambda_\circ$, which parametrizes the vertex correction,
 is not small. It is a maximum for ${\bf k=k_h}$, where ${\bf k_h}$ is a ``hot
 spot'' on the Fermi surface. This makes large quantitative corrections to the
 theory.     
\end{abstract}
\vspace{.5cm}
\pacs{PACS Numbers:   }

Most theories of the cuprate superconductors are either (i) phenomenological attempts to tie together many different experiments, or (ii) attempts at a ``microscopic" derivation, beginning from models such as the Hubbard or t-J Hamiltonians.

Of the variety of phenomenological theories, one of the most widely-discussed is that involving antiferromagnetic (AFM) spin fluctuations in a random-phase approximation (RPA) model of a nearly AFM Fermi liquid, in the optimally-doped regime \cite{1,2,3}. For weak doping a Non-Linear Sigma (NL$\sigma$) model is used \cite{4,5,6}, but no attempt is made to deal with the intervening metal-insulator transition. This transition plays a central role in the more microscopic theories, many of which do not yield a Fermi liquid state except at high T and/or rather large doping; instead one finds that ``singular interactions" are generated \cite{7,8,9}, and these inevitably give a non-Fermi liquid normal state \cite{7,8,9,10}, even at optimal doping. However they also yield strong AFM spin fluctuations \cite{10,11}, as do some other microscopic approaches; a wide variety of phenomenological approaches involving AFM spin fluctuation is possible.

To decide between competing theories it is useful to check their internal consistency. Several analyses have been made of the critical properties of AFM spin fluctuations \cite{11,12,13}, but here we wish to focus specifically on the Fermi liquid model \cite{1,2,3}. We find, in agreement with earlier work \cite{12,13}, that once the nearly AFM form of the spin fluctuation spectrum,
\beq
\chi ({\bf q},\omega)={\chi_Q\over {1+\xi^2({\bf q-Q})^2-i\omega /\omega_{SF}}}
\label{eq.1}
\eeq
has been assumed for ${\bf q}$ near ${\bf Q}=(\pm{\pi\over a},\pm{\pi\over a})$,
and provided we assume from the outset a Fermi liquid model,
 then no singularities strong enough to destroy Fermi liquid behavior emerge in
 the fermion spectrum, provided $\xi$ and $\omega_{SF}$ are finite. However we
 also find that a detailed quantitative determination of vertex corrections
 shows that previous estimates of these have been over-optimistic - they are
 not small. This has important quantitative repercussions for the theory. 

\begin{figure}
\epsfysize 18cm
\epsfbox[-10 150 550 800]{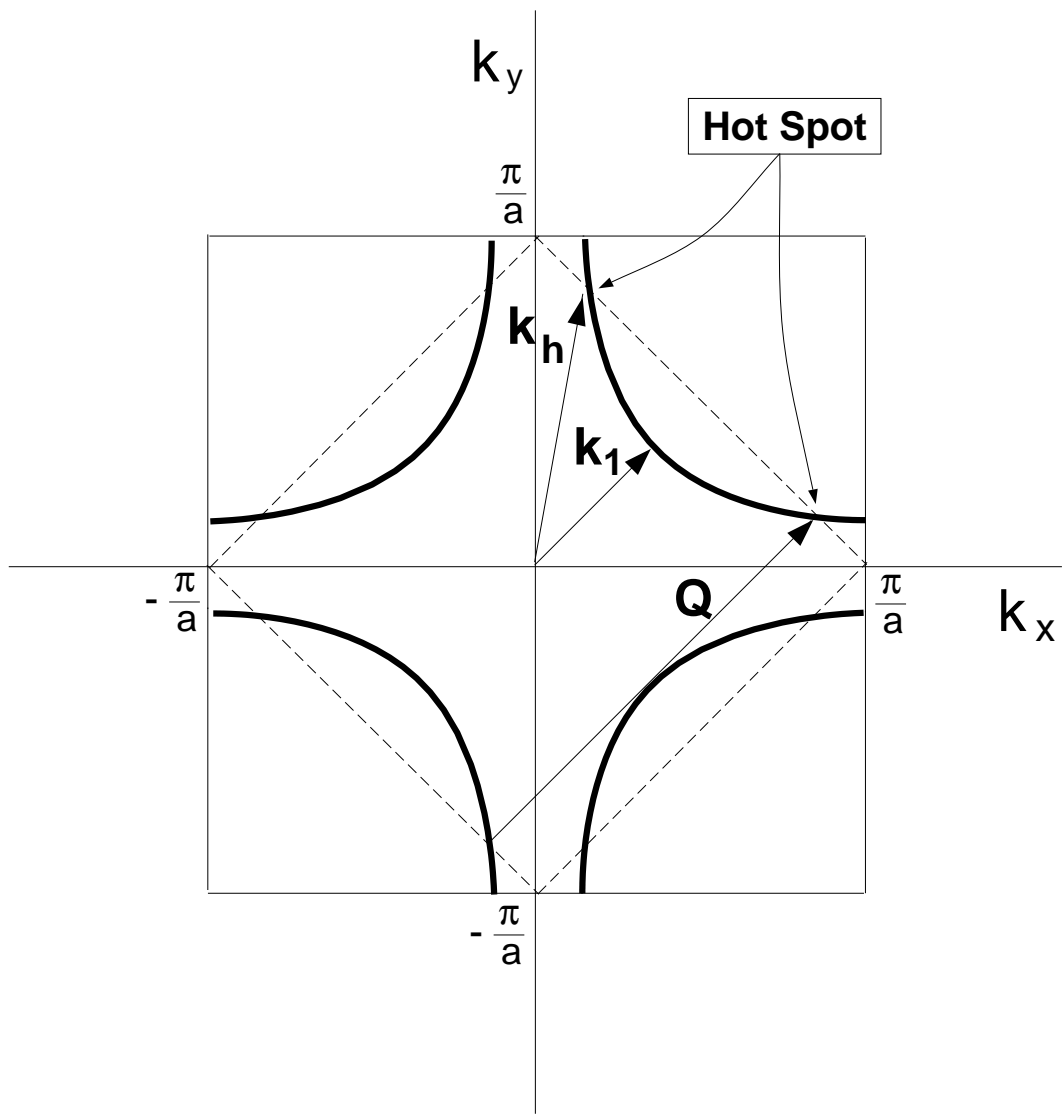}
\caption[Fermi surface in the first Brillouin zone]{Fermi surface in the first Brillouin zone, with the value for $t$ and $t'$ given in the text; and we assume n=0.75. Calculations are presented here for the wave vectors $k_1$ and $k_h$.}
\label{fig1}
\end{figure}

\begin{figure}
\epsfysize 8cm
\epsfbox[40 260 468 500]{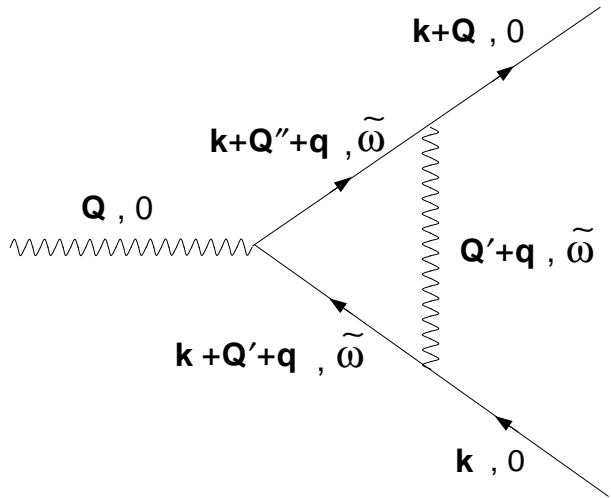}
\caption{First correction $\delta\Lambda_{\bf k}$ to the bare vertex, for an incoming fermion with momentum {\bf k} and energy 0 (relative to the Fermi energy), interacting with a fluctuation of wave-vector {\bf Q} and zero energy.\label{fig2}}
\end{figure}

We start from a model Hamiltonian
\beq
{\cal H}= \sum_{{\bf p},\sigma}\epsilon_{\bf p}\psi_{{\bf p}\sigma}^{\dag}\psi_{{\bf p}\sigma}+
{\bar{g}\over 2}\sum_{{\bf q,k,}\alpha,\beta}\psi_{{\bf k+q}\alpha}^{\dag}\psi_{{\bf k}\beta}
{\bf {\bf \sigma}_{\alpha\beta}.S}(-{\bf q})
\label{eq.2}
\eeq
where
\beq
\epsilon_{\bf p}=-2t[\cos(p_xa)+\cos(p_ya)]-4t^{\prime} \cos(p_xa)\cos(p_ya)
-\mu
\label{eq.3}
\eeq
is the quasiparticle dispersion relation \cite{3,14}, with $t=0.25$eV and $t^{\prime}=-0.45 t$. The presence of $t^{\prime}$ allows ``hot spots" on the Fermi surface $S_F$ (FIG. \ref{fig1}) which can be connected by ${\bf Q}$; this leads to singular behavior when the gap in (\ref{eq.1}) disappears \cite{13,15}. ${\bf S}({\bf r})$ is the spin fluctuation density operator; use of (\ref{eq.2}), with a bosonic spin fluctuation propagator (\ref{eq.1}) and a phenomenological interaction $\bar{g}$, allows one to develop the usual ``paramagnon" effective field theory \cite{16}. The first vertex correction can be calculated for general external momenta \cite{17}; here we concentrate on the interaction between a Fermi surface electron and a spin fluctuation with ${\bf q}={\bf Q}$ (FIG. \ref{fig2}) for which the lowest vertex correction is

\beq
{\delta\Lambda_{\bf k}\over \Lambda_\circ} = -{{\bar g}^2\over 4}V\sum_{{\bf Q^\prime}}\int {d^2q\over (2\pi)^2}{d{\tilde \omega}\over 2\pi}\chi ({\bf Q^{\prime}+q},{\tilde \omega})G({\bf k+Q^{\prime}+q},{\tilde \omega})G({\bf k+Q^{\prime\prime}+q},{\tilde \omega})
\label{eq.4}
\eeq
where $\delta\Lambda_{\bf k}=\delta\Lambda({\bf k},\epsilon=0;{\bf q=Q},\Omega=0)$ and ${\bf Q^{\prime\prime}}={\bf Q}+{\bf Q^{\prime}}$; the sum over 
${\bf Q^{\prime}}$ takes care of Umklapp processes.
In the reduced Brillouin zone ${\bf Q^{\prime\prime}}=(0,0)$.
Note that the overall sign of the graph in FIG. \ref{fig2} is 
{\it negative}, ie., eqtn. (\ref{eq.4}) is negative; the bare vertex 
$\Lambda_{\circ} = {\bar {g}\over 2}$ in eqtn. (\ref{eq.2}) is positive.

We concentrate on $\delta\Lambda_{\bf k}$, with ${\bf k}$ $\epsilon$ $S_F$ 
(here $S_F$ is the Fermi surface) because it gives an indication of the size 
of vertex corrections. This is because in general 
$|\delta\Lambda({\bf k,\epsilon;q},\Omega)|$ exceeds   $|\delta\Lambda_{\bf k}|$ 
(in fact it diverges along a surface in k-space, for given values of 
${\bf q},\Omega$ and $\epsilon$); thus one cannot argue, even after integrating 
over one or more of its arguments, that 
$|\delta\Lambda({\bf k,\epsilon;q},\Omega)|$ will lead to corrections smaller 
than one would get from just using $\delta\Lambda_{\bf k}$ (our argument here 
parallels Migdal's \cite{24}). More generally one finds that if 
$\Omega\ll{\rm qv_F},\Delta$ , where $\Delta$ is the spin gap, then 
$\delta\Lambda_{\bf k}$ is a good approximation to 
$\delta\Lambda({\bf k,\epsilon;q},\Omega)$. Writing (\ref{eq.4}) as

\begin{eqnarray}
{\delta\Lambda_{\bf k}\over \Lambda_\circ} &=&-{{\bar g}^2\over 4}V\sum_{{\bf Q^\prime}}\int {d^2q\over (2\pi)^2}{d\omega\over \pi}{\chi^{\prime\prime} ({\bf Q'+q},\omega)\over {\epsilon_{\bf k+Q^{\prime}+q}-\epsilon_{\bf k+Q^{\prime\prime}+q}}}({f_{\bf k+Q^{\prime}+q}\over {\epsilon_{\bf k+Q^{\prime}+q}-\omega}}+ {{1-f_{\bf k+Q^{\prime}+q}}\over {\epsilon_{\bf k+Q^{\prime}+q}+\omega}}\nonumber\\&-&{f_{\bf k+Q^{\prime\prime}+q}\over {\epsilon_{\bf k+Q^{\prime\prime}+q}-\omega}}-{{1-f_{\bf k+Q^{\prime}+q}}\over {\epsilon_{\bf k+Q^{\prime}+q}+\omega}}) = -g^2{\chi_Q\over {12\pi^3\omega_{SF}}} ({\omega_{SF}\over \mu})^2{\cal I_{\bf k}}
\label{eq.5}
\end{eqnarray}
where  $g^2={3\over 4}{\bar g}^2$ and ${\cal I_{\bf k}}$ is dimensionless; we
define $g$ to correspond directly with the coupling constant $g$ used in
Monthoux and Pines \cite{3}.
At zero temperature one has
\beq
{\cal I_{\bf k}} = \int_{-2\pi}^{2\pi}d{\bar q}_x \int_{-2\pi}^{2\pi}d{\bar q}_y {1\over {{\bar \epsilon}_1-{\bar \epsilon}_2}}[{\rm Sgn}({\bar \epsilon}_1)G_1({\bf {\bar q}},{\bar \epsilon}_1)- {\rm Sgn}({\bar \epsilon}_2)G_1({\bf {\bar q}},{\bar \epsilon}_2)
-G_2({\bf {\bar q}},{\bar \epsilon}_1)+G_2({\bf {\bar q}},{\bar \epsilon}_2)]
\label{eq.6}
\eeq
where ${\bar q_x} = q_xa$, etc.; ${\bar \epsilon}_1=\epsilon_{\bf k+Q^{'}+q}/\mu$ and ${\bar \epsilon}_2=\epsilon_{\bf k+Q^{''}+q}/\mu=\epsilon_{\bf k+q}/\mu$. $G_1$ and $G_2$ are defined by
\beq
G_1({\bf q},{\bar \epsilon})={\pi X\over 2({\bar \epsilon}^2+X^2)}
\label{eq.7}
\eeq
\beq
G_2({\bf q},{\bar \epsilon})={{\bar \epsilon} \ln (X/|{\bar \epsilon}|)\over {\bar \epsilon}^2+X^2}
\label{eq.8}
\eeq
with $X=[1+({\xi\over a})^2({\bar q}_x^2+{\bar q}_y^2)](\omega_{SF}/|\mu|)$.

${\cal I_{\bf k}}$ can be investigated both analytically and numerically. Here we calculate it for two different points in k-space, both on the Fermi surface (see FIG. \ref{fig1}); ${\bf k_1}$ makes a $45^o$ angle with $k_x$, and ${\bf k_h}$ is a ``hot spot" wave vector. 

In order to obtain a value for $\delta\Lambda_{\bf k}$, we need values for 
the spin fluctuation energy $\omega_{SF}$, the correlation length $\xi$, 
the susceptibility $\chi_Q$, the coupling constant $g$, and the chemical 
potential $\mu$ (which is determined by the electron filling factor n). 
There are different values reported in the references \cite {3,18,19}. We 
have evaluated $\delta\Lambda_{\bf k}$ for ${\bf k}={\bf k_1},{\bf k_h}$, 
in two ways, viz. (a) by assuming various published values for the 
different parameters, and (b) by making the simple assumption that 
one is very close to
an AFM instability, and then, in the spirit of RPA, imposing the 
condition  $|({\bar g}/2)\Pi^{(1)} ({\bf Q},0)|=1$ where 
$\Pi^{(1)} ({\bf Q},\omega)$ is the electron-hole bubble. Numerical 
calculation gives  $|\Pi^{(1)} ({\bf Q},0)|= 2.6 ({\rm eV})^{-1}$ and 
thereby $g=0.67$ eV, assuming n=0.75 and the band structure in (\ref{eq.3}). 
Since $\omega_{SF}/|\mu| \ll 1$, this value of $g$ should be a very 
good guess, within a naive RPA scheme. 

The results are summarized in Table \ref{table1}. We use two different values 
for $\omega_{SF}$ ; these are the two different values quoted by 
Monthoux and Pines {\it et al}. \cite{3}. We also use two different values for 
the coupling constant $g$, quoted from MPI and MPII (see ref. \cite{3} 
again). We use values of $\xi=2.5a$, $\chi_Q = 80$ states/eV (from \cite{3}) 
and n=0.75 , appropriate to  ${\rm YBa_2Cu_3O_7}$ (again quoted 
from \cite{3}). This value of n corresponds, with the band structure in 
(\ref{eq.2}), to a value of $|\mu|\sim 1.46t \equiv 0.365$ eV.   

We see that even the values for the vertex correction calculated from the simple RPA model (b) are not small; as in the standard discussion of Migdal's theorem, the importance of vertex corrections appears in the ratio $|\delta\Lambda_{\bf k}|/\Lambda_\circ$. If one takes values of $g$ from the literature \cite{3}, this ratio is quite unreasonably large (as large as 2.43 for the hot spots in the model used by MPI). Thus vertex corrections are clearly very important. The values we quote for $|\delta\Lambda_{\bf k}|/\Lambda_\circ$ are considerably larger than previous estimates \cite{3,21,22}.  The reason for this difference with previous work can be tracked back to the factor ${\cal I}_{\bf k}$, which is impossible to guess from purely dimensional arguments. In fact if we drop the factor ${\cal I}_{\bf k}$ from $\delta\Lambda_{\bf k}$, we get an order of magnitude estimate for $\delta\Lambda_{\bf k}$ given by
\beq
{|\delta\Lambda_{\bf k}|\over \Lambda_\circ} \sim O[g^2{\chi_Q \over 4\pi^3|\mu|}{\omega_{SF} \over |\mu|}] \ll 1\nonumber
\label{eq.9}
\eeq
which is broadly in agreement with previous estimates (see eg. Millis \cite{21})
; $\chi_Q$, $\omega_{SF}$ and $g$ must be redefined to conform with the parametrizations in this paper).

In fact however  ${\cal I}_{\bf k}$ is surprisingly large, and also shows 
a significant variation around the Fermi surface, with a maximum at the hot spots, and a minimum at intermediate wave-vector like ${\bf k_1}$.  We should emphasize here that analytic calculations of $\delta\Lambda_{\bf k}$ 
have to be approximated rather carefully in order to give reasonable agreement with the numerical results in Table \ref{table1}. Approximations such as those of Hertz {\it et al.} \cite{23} (see also \cite{22}), which try to separate off a rapidly-varying (in {\bf q}-space) contribution from $\chi'' ({\bf q},\omega)$, give quantitatively incorrect results (including a completely unphysical $\ln [{\bf (k-k_h)}a]$ divergence as one approaches the hot spot). A more detailed discussion of the behavior of  ${\cal I}_{\bf k}$ is given in ref. \cite{17}.

One might suppose that ${\cal I}_{\bf k}$ is large simply because of the
band structure (ie., because of van Hove singularities, or the hot spots).
If this were true one could argue that the quasiparticle weight ought to be
renormalised down near these singular points in the Brillouin zone, and that
this would considerably reduce the vertex correction. In fact however we find 
this is not the case; this can be checked analytically by suppressing the 
regions immediately around the hot spots in the integral for 
${\cal I}_{\bf k}$, or by simply redoing the numerical calculation for
a slightly different band structure. We find that suppressing the hot spots 
entirely, reduces the vertex correction by a factor which is everywhere less
than 2 (and which differs very little from unity when {\bf k}
is far from a hot spot). Thus we do not believe that incorporating self-
energy corrections near the hot spots would significantly reduce the vertex
correction. 

We re-emphasize here that these results do not
necessarily invalidate the internal consistency
of the Fermi liquid starting point, in this theory. However they do show that
the theory cannot be trusted quantitatively, at least in the usual RPA form.
As is well known the RPA is not a ``conserving approximation", and for spin
fluctuation theories this makes it unreliable
 (cf. ref.\cite{16}, especially
section 3). It is useful to compare the case 
of nearly ferromagnetic $^3$He liquid,
where vertex corrections are also quite large, and where use of the paramagnon
model yields values for $m^*/m$ which are off by a large factor 
\cite{20}. Thus if we use melting curve Landau parameters,
$F_0^A \sim 0.75$ and $F_1^S \sim 15$, we infer a value for the Stoner factor
$S \sim 24$ which yields $m^*/m = {9\over 2} \ln S \sim 15$ , in the paramagnon
model. This is 
roughly 2.5 times the correct value of $\sim 6$ (note that the
first vertex correction is $\delta\Lambda/{\bar I} \sim \ln S \sim 3$
in this model), and no amount of self-consistent summing of
diagrams can cure this numerical problem.

Similar problems can clearly occur in the present AFM spin fluctuation model.
We believe this is the main reason for the
difficulty one encounters in the MP models, in determining a value for $g$
that (a) gives the correct superconducting $T_c$, and (b) is consistent with
 the observed spin susceptibility. 

To check the structure at higher order, we have also estimated 
the contributions from the graphs
containing 2 spin fluctuation lines (there are actually 7 distinct
 graphs at this
level), and found that some of them are also large for the values of $g$ used
above \cite{17}. Thus,
just as for the case of nearly ferromagnetic
$^3$He,  we see no reason to believe, 
{\it for the values of the parameters given in the
table}, that performing infinite graphical sums 
will lead to results which 
are numerically more reliable, even if they do converge to some smaller 
renormalised vertex-there will always be other diagrams with large values,
which will in general give uncontrolled contributions. 

It is interesting to compare these results with some other recent
investigations. In the weak-coupling limit, 
Chubukov \cite{13} has calculated the leading vertex corrections to $g$, 
concentrating on the gapless case; in the case where is a gap, he finds that
the renormalised coupling is also small (for realistic values of the gap).
On the other hand Schrieffer\cite{25} has argued that a correct
formulation of the theory, even 
in the weak-coupling limit, must take account of the short-range local 
antiferromagnetic order even in the normal state-if this done, he finds
that a weak-coupling calculation shows very strong
{\it suppression} of the vertex
when one is close to the antiferromagnetic transition. This theory seems
rather interesting-note that a related calculation by Vilk and Tremblay
\cite{26}
finds that the existence of such a 
short-range antiferromagnetic order will cause a breakdown
of the Fermi-liquid starting point itself! Thus 
the question of what is the correct theory itself is rather
confused, even in the weak-coupling regime. It is certainly not clear how any of
these arguments will work in the regime discussed in this paper, when $g$ is 
not small enough to control the magnitude of the vertex corrections. 

It is of course crucial that these higher-order corrections also be
included in any version of this theory that tries to reconcile different
experiments - as emphasized by Pines \cite{19}, the 
justification of the theory stands or falls on
its ability to do this 
{\it quantitatively}. It is possible that such a programme
might succeed if one can show that the actual parameters $g$, $\omega_{SF}$,
and $\chi_Q$ are such that $|\delta\Lambda_{\bf k}|/\Lambda_\circ$ is considerably
less than one
(ie., if one is genuinely in the weak-coupling regime).
 This would also be true of  
versions of the theory in which $\omega_{SF}$ depends on $g$,
whilst the spin gap becomes an independent parameter\cite{13}; or
of the theory of
Schrieffer cited above\cite{25}.
On the other hand if  $|\delta\Lambda_{\bf k}|/\Lambda_\circ > O(1)$, we see no
hope that such a scheme could succeed 
quantitatively (in., eg., the calculation of $T_c$),
 since the vertex corrections
become large.

We would like to thank I.Affleck, V.Barzykin, D.Bonn, and W.Hardy for useful
conversations, as well as A.Chubukov 
for correspondence.
This work was supported by NSERC in Canada, and by the CIAR.
\begin{table}
\begin{center}
\begin{tabular}{|c|c|c|c|c|c|c|}  \hline
              & $g$ (eV) & $\omega_{SF}$ (meV) & ${\cal I}_{\bf k_1}$ & $\delta
\Lambda_{\bf k_1}/\Lambda_\circ$ & ${\cal I}_{\bf k_h}$ &
 $\delta\Lambda_{\bf k_h}/\Lambda_\circ$\\ \hline\hline
 
MPI        & 1.36 & 7.7 & 78.6 & -1.81 & 105.6 & -2.43 \\ \hline
MPII       & 0.64 & 14 & 49.6 & -0.46 & 73.4 & -0.68 \\ \hline 
``RPA"     & 0.67 & 7.7 & 78.6 & -0.44 & 105.6 & -0.59 \\ \cline{3-7}
           &      & 14 & 49.6 & -0.50 & 73.4 & -0.74 \\ \hline
\end{tabular}
\end{center}
\caption[Calculated values of the vertex correction $\delta\Lambda_{\bf k}$ for two different wave-vectors $\bf k_1$ and $\bf k_h$ on the Fermi surface.]
{Calculated values of the vertex correction $\delta\Lambda_{\bf k}$ for two different wave-vectors $\bf k_1$ and $\bf k_h$ on the Fermi surface (columns 4 and 6 in the table). The papers MPI and MPII (ref. \cite{3}) give different values for $g$, and different values for $\omega_{SF}$. From the values for these two models one calculates ${\cal I}_{\bf k}$ in equation (\ref{eq.6}), and thence $\delta\Lambda_{\bf k}/\Lambda_\circ$. The third model is the naive ``RPA" model described in the text, for which $g$ is determined; we have calculated  ${\cal I}_{\bf k}$ and $\delta\Lambda_{\bf k}/\Lambda_\circ$ for two values of $\omega_{SF}$ given in MPI and MPII respectively.}
\label{table1}
\end{table}

\begin{references}
\bibitem{1} D.J.Scalapino, to be published
\bibitem{2} T.Moriya, Y.Takahashi, K.Ueda, J. Phys. Soc. Jap. {\bf 59}, 2905 (1990), and J. Phys. Chem. Sol. {\bf 53}, 1515 (1992) 
\bibitem{3} A.J.Millis, H.Monien, D. Pines, Phys. Rev. B {\bf 42}, 167 (1990); and P. Monthoux, D. Pines, Phys. Rev. B {\bf 47}, 6069 (1993) [MPI]  and  Phys. Rev. B {\bf 49}, 4261 (1994) [MPII]
\bibitem{4} S.Chakravarty, B.I.Halperin, D.Nelson, Phys. Rev. B {\bf 39}, 2344 (1989)
\bibitem{5} A.V. Chubukov, S. Sachdev, T. Senthil, Nucl. Phys. B {\bf 426}, 601, (1994), and references therein.
\bibitem{6} A.Sokol, D.Pines, Phys. Rev. Lett. {\bf 71}, 2813 (1993)
\bibitem{7} P.W.Anderson, Phys. Rev. Lett. {\bf 65}, 2306 (1990)
\bibitem{8} P.C.E.Stamp, Phys. Rev. Lett. {\bf 68}, 2180 (1992), and  J. de Physique {\bf I, 3}, 624 (1993)
\bibitem{9} L.Ioffe, A.Larkin, Phys. Rev. B {\bf 39}, 8988 (1989); P.A.Lee, N.Nagaosa, Phys. Rev. B {\bf 46}, 5621 (1992)
\bibitem{10} D.V.Khveschenko, P.C.E.Stamp, Phys. Rev. Lett. {\bf 71}, 2118 (1993), and Phys. Rev. B {\bf 49}, 5227 (1994); J.Gan, E.Wong, Phys. Rev. Lett. {\bf 71}, 4226 (1993); B.L.Altshuler, L.B.Ioffe, A.J.Millis, Phys. Rev. B {\bf 50}, 14048 (1994); S.Chakraverty, R.E.Norton, O.V.Syljuasen, Phys. Rev. Lett. {\bf 74}, 1423 (1995); C.Nayak, F.Wilczek, Nucl. Phys. B {\bf 430}, 534 (1994); and references therein.

\bibitem{11} B.L.Altshuler, L.Ioffe, A.I.Larkin, A.J.Millis, preprint
\bibitem{12} A.J.Millis, Phys. Rev. B {\bf 48}, 7183 (1993)
\bibitem{13} A.V.Chubukov, Phys. Rev. B {\bf 52}, R3840 (1995)
\bibitem{14} J.Yu, S.Massidda, A.J.Freeman, Phys. Lett. A {\bf 122}, 198 (1987); Q.Si, Y.Zha, K.Levin, J.P.Lu, Phys. Rev. B {\bf 47}, 9055 (1993)
\bibitem{15} R.Hlubina, T.M.Rice, Preprint
\bibitem{16} P.C.E.Stamp, J. Phys. F {\bf 15}, 1827 (1985), and references therein.
\bibitem{17} M.H.Sharifzadeh Amin, P.C.E.Stamp, Longer paper in preparation
\bibitem{24} A.B.Migdal, Sov. Phys. JETP {\bf 7}, 996 (1958)
\bibitem{18} P. Monthoux, A. Balatsky, D. Pines, Phys. Rev. B {\bf 46}, 14803 (1992)
\bibitem{19} D. Pines, in " High $T_c$ Superconductivity and the $C_{60}$ Family ", ed. T.D.Lee, H.C.Ren (Gordon and Breach, 1995).
\bibitem{20} See in this connection W.F.Brinkman, in Proc. 24th Nobel Symposium (Stockholm, Nobel Foundation) 1973.
\bibitem{21} A.J.Millis, Phys. Rev. B {\bf 45}, 13047 (1992)
\bibitem{22} I.Grosu, M.Crisan, Phys. Rev. B {\bf 49}, 1296 (1994). Our results do not agree with this paper. The form of equations (1) and (5) of Grosu and Crisan seem to be appropriate to a nearly ferromagnetic system - their expansion are being made near ${\bf q}=0$ (rather than ${\bf q} \sim {\bf Q}$).
\bibitem{23} J.A.Hertz, K.Levin, M.T Beal-Monod, Phys. Solid State Com. {\bf 18}, 803 (1976)
\bibitem{25} J.R.Schrieffer, J. Low Temp. Phys. {\bf 99}, 397 (1995). 
Note that the {\it sign} of our correction is negative, which one might
consider to be at least consistent with Schrieffer's result-however this
could be mere coincidence, since some of the higher corrections have
opposite sign, and in any case the {\it magnitude} of our correction is
so large that it is impossible to say what the corrections might sum
to.
\bibitem{26} Y.M.Vilk, A.-M.S. Tremblay, Europhys. Lett. {\bf 33}, 159 (1996).
\end{references}
\end{document}